\documentclass[12pt]{article}
\usepackage{graphics}
\begin{document}
\section*{Microscopic chaos from Brownian motion?}

In a recent Letter in {\em Nature}, Gaspard et al.~\cite{G} claimed to
present empirical evidence
for microscopic chaos on a molecular scale from an ingenious experiment
using a time series of the positions of a Brownian particle in a liquid.
The Letter was preceded by a lead
article~\cite{DS} emphasising the fundamental nature of the experiment.
In this note we demonstrate that virtually identical results 
can be obtained by analysing a corresponding numerical time series of a
particle in a manifestly microscopically nonchaotic system.

As in Ref.~\cite{G} we analyse the position of a single particle
colliding with many others. We use the Ehrenfest wind-tree model~\cite{E}
where the pointlike (``wind'') particle moves in a plane colliding with
randomly placed fixed square scatterers (``trees'', Fig. 1a). We choose
this model because collisions with the flat sides of the squares do not
lead to exponential separation of corresponding points on initially nearby
trajectories.  This means there are no positive Lyapunov exponents which
are characteristic of microscopic chaos here. In contrast the Lorentz model
used in~\cite{G} as an example similar to Brownian motion is a wind-tree
model where the squares are replaced by hard (circular) disks (cf.\cite{G},
Fig.~1) and exhibits exponential separation of nearby trajectories, leading
to a positive Lyapunov exponent and hence microscopic chaos.

Nevertheless, we now demonstrate that the nonchaotic Ehrenfest model
reproduces all the results of Ref.~\cite{G}.  A particle trajectory
segment shown in Fig.~1b is strikingly similar to that for the Brownian
particle, (cf.\cite{G}, Fig.~2).  Our subsequent analysis parallels that of
Ref.~\cite{G}, where more details may be found.  Thus the microscopic
chaoticity is determined by estimating the Kolmogorov-Sinai entropy
$h_{KS}$, using the method of Procaccia and others~\cite{GP,CP}
via the information entropy $K(n,\epsilon,\tau)$
obtained from the frequency with which the partical retraces part of
its (previous) trajectory within a distance $\epsilon$ for $n$
measurements spaced at a time interval $\tau$.  Since for the systems
considered here $h_{KS}$ equals the sum of the positive Lyapunov exponents,
the determination of a positive $h_{KS}$ would imply microscopic chaos.
As in~\cite{G} we find that $K$ grows linearly with time
(Fig 1c and~\cite{G} Fig.~3), giving a positive (non-zero) bound on $h_{KS}$
(Fig 1d and~\cite{G} Fig.~4).  Indeed our Figs.~1b-d for a microscopically
nonchaotic model are virtually identical with the corresponding figures
2-4 of~\cite{G}.  Therefore Gaspard et al. did not prove microscopic
chaos for Brownian motion.

The algorithm of~\cite{GP,CP} as applied here cannot determine the
microscopic chaoticity of Brownian motion since the time interval
between measurements, $1/60\;s$ in~\cite{G}, is so much larger than
the microscopic time scale determined by the inverse collision frequency
in a liquid, approximately $10^{-12}\;s$.  A decisive determination of
microscopic chaos would involve, it seems at the very least, a time
interval $\tau$ of the same order as characteristic microscopic time scales.

\noindent
\\
{\bf C. P. Dettmann, E. G. D. Cohen}\\
Center for Studies in Physics and Biology,\\
Rockefeller University,\\
New York, NY 10021, USA\\
\\
{\bf H. van Beijeren}\\
Institute for Theoretical Physics,\\
University of Utrecht,\\
3584 CC Utrecht, The Netherlands

%\newpage

%\newpage
\pagestyle{empty}
\begin{figure}[h]
\begin{picture}(350,360)
\put(-50,-100){\scalebox{.8}{\includegraphics{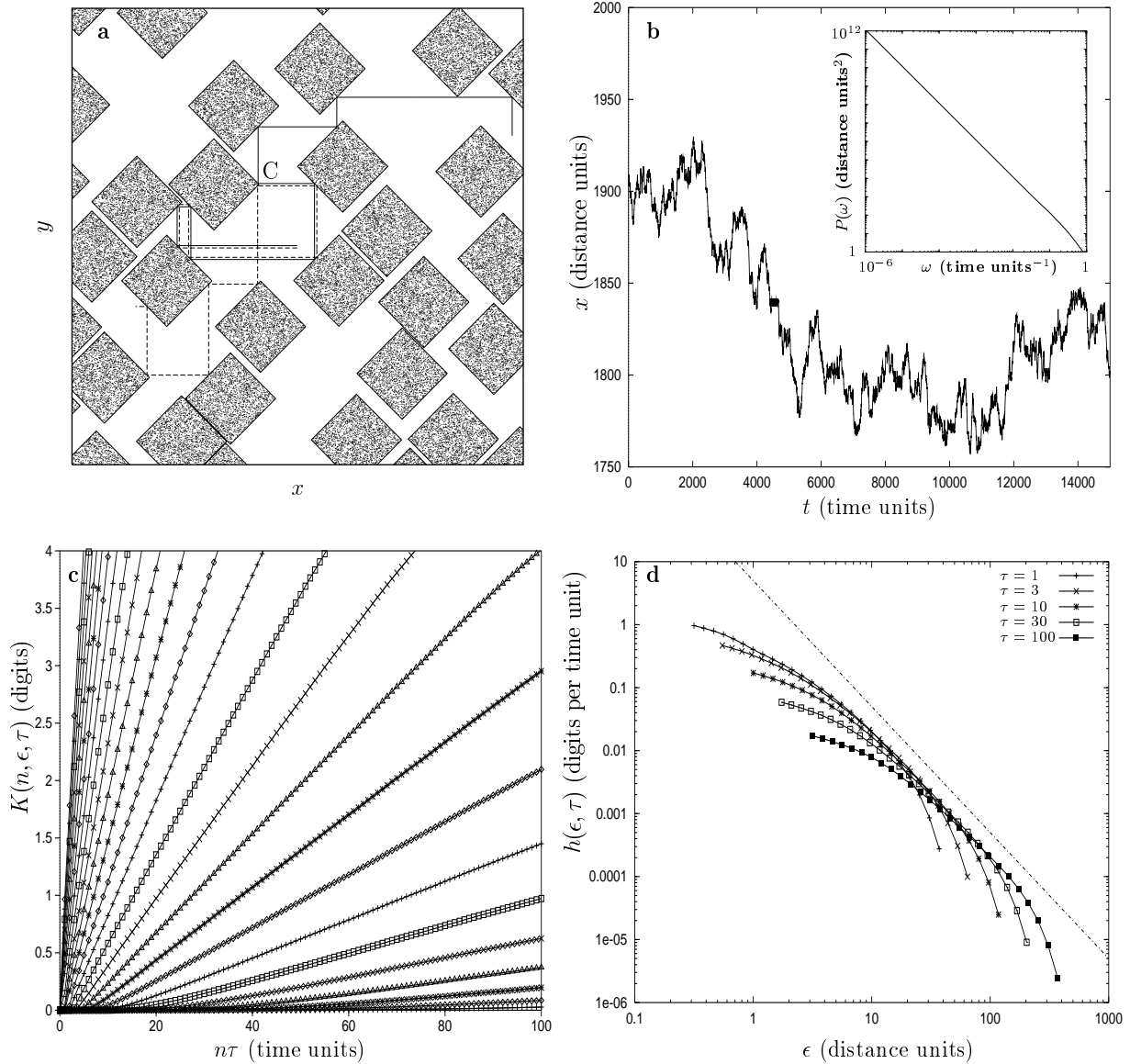}}}
\end{picture}
\caption{Brownian motion results of~\protect\cite{G}
numerically reproduced from the nonchaotic Ehrenfest wind-tree model
(notation as in~\cite{G}). The square scatterers have a diagonal of
$2$ length units and fill half the area considered.  The particle moves
with unit velocity in four possible directions.  The position on its
trajectory is determined for $10^6$ points separated by one time unit.
({\bf a}) Two nearby trajectories split only at a corner C; no exponential
separation occurs (cf~\cite{G}, Fig.~1).  ({\bf b}) A typical trajectory is
diffusive, with an $\omega^{-2}$ power spectrum (inset), cf~\cite{G},
Fig.~2.  ({\bf c}) The information entropy $K(n,\epsilon,\tau)$
for $\tau=1$ and $\epsilon=0.316\times 1.21^m$ with $m=0\ldots25$,
cf~\cite{G}, Fig.~3.  ({\bf d}) The envelope of the slopes of these
$K$-curves, $h(\epsilon,\tau)$ appears to imply a positive, ie. chaotic,
$h_{KS}$ for the Ehrenfest model, as for Brownian
motion (cf.~\cite{G}, Fig.~4).}
\end{figure}

\end{document}